\documentclass[conference]{IEEEtran} 

\usepackage{type1cm}    
\usepackage{enumerate}
\usepackage{graphicx}
\usepackage{epstopdf}
\usepackage{epsfig}
\usepackage{color}
\usepackage{multirow}
\usepackage{etoolbox}
\usepackage{booktabs}
\usepackage{amsmath}
\usepackage{amssymb}
\usepackage{mathtools}
\usepackage{hhline}
\usepackage{setspace}
\usepackage{balance}
\usepackage{xcolor,colortbl}
\usepackage{array}
\usepackage{xspace}     
\usepackage{balance}     
\usepackage{caption}
\usepackage{siunitx}    
\usepackage[vlined,linesnumbered,ruled,noend]{algorithm2e}
\usepackage{lscape}
\usepackage{subfigure}
\usepackage{array}
\usepackage{pifont}
\usepackage{alltt}
\usepackage{listings}
\usepackage{multirow}
\usepackage{hyperref}
\usepackage{makecell}

\usepackage{url}

\usepackage{breakurl}

\definecolor{Gray}{gray}{0.90}
\definecolor{persiangreen}{rgb}{0.0, 0.65, 0.58}
\newcommand{\cmark}{\color{persiangreen}\ding{51}}%

\newcolumntype{a}{>{\columncolor{Gray}} p{1.7cm}}

\newcommand{\js}{JavaScript\xspace}
\newcommand{\name}{MarioNet\xspace}
\begin{document}
\title{Master of Web Puppets: Abusing Web Browsers\\for Persistent and Stealthy
Computation}

\author{\makebox[.99\linewidth]{
		Panagiotis Papadopoulos,\IEEEauthorrefmark{1}
		Panagiotis Ilia,\IEEEauthorrefmark{1}
		Michalis Polychronakis,\IEEEauthorrefmark{2}}
		\and \makebox[.99\linewidth]{
		Evangelos P. Markatos,\IEEEauthorrefmark{1}
		Sotiris Ioannidis,\IEEEauthorrefmark{1}
		Giorgos Vasiliadis\IEEEauthorrefmark{1}}\\
	
	\and \makebox[.01\linewidth]{} 
	\and
	
	\IEEEauthorblockA{\IEEEauthorrefmark{1}
		FORTH, Greece\\ 
		\{panpap, pilia, markatos, sotiris, gvasil\}@ics.forth.gr}
		
	\and
	\IEEEauthorblockA{\IEEEauthorrefmark{2}
		Stony Brook University, USA\\
		mikepo@stonybrook.edu}
}

\maketitle

\begin{abstract}
  The proliferation of web applications has essentially transformed
modern browsers into small but powerful operating systems.
Upon visiting a website, user devices run implicitly trusted script code,
the execution of which is confined within the browser
to prevent any interference with the user's system.
Recent \js APIs, however, provide advanced capabilities that not only
enable feature-rich web applications, but also allow attackers to
perform malicious operations despite the confined nature of \js
code execution.

In this paper, we demonstrate
the powerful capabilities that modern browser APIs provide to attackers
by presenting \name: a framework that allows a remote malicious entity to 
control a visitor's browser and abuse its resources for
unwanted computation or harmful operations, such as cryptocurrency mining,
password-cracking, and DDoS.
\name\ relies solely on already available HTML5 APIs,
without requiring the installation of any additional software.
In contrast to previous browser-based botnets,
the persistence and stealthiness characteristics of \name
allow the malicious computations to continue in the background
of the browser even after the user closes the window or tab of the initial malicious website.
We present the design, implementation, and evaluation of a prototype
system, \name, that is compatible with all major browsers,
and discuss potential defense strategies to counter the threat
of such persistent in-browser attacks.
Our main goal is to raise awareness regarding this new class of attacks,
and inform the design of future browser APIs
so that they provide a more secure client-side environment for web
applications.

\end{abstract}

\section{Introduction}
\label{sec:introduction}

Our increasing reliance on the web has 
resulted in sophisticated browsing software that essentially
behaves as an integrated operating system for web applications.
Indeed, contemporary browsers provide
an abundance of APIs and sensors (e.g., gyroscope, location, 
battery status) that can be easily used by web applications
through locally-running \js code.
The constantly expanding \js interfaces available in modern browsers
enable users to receive timely updates, render interactive maps
and 3D graphics, or even directly connect to other
browsers for peer-to-peer audio or video communication (e.g., through WebRTC).

In the era of edge computing, these capabilities have pushed a significant
part of web application logic to the endpoints. Web publishers transfer 
parts of the critical computations on the user side, thus minimizing 
latency, providing satisfactory user experience and usability, while at 
the same time increasing the scalability of the provided service. 
Despite all these advancements, the web largely works in the very 
same way since its initial inception: whenever a user visits a website, the 
browser requests from the remote web server (and typically from other
third-party servers) all the necessary components (e.g., HTML, CCS,
\js, and image files), 
executes any script code received, and renders the website 
locally. Thus, whenever a user browses to a website, the browser 
blindly executes any received \js code on the user's machine.

From a security perspective,
a fundamental problem of web applications is that by default their publisher is
considered as trusted, and thus allowed to run \js code (even
from third parties) on the user side without any restrictions (as long as it is
allowed by the site's content security policy, if any).
More importantly, users remain oblivious about the
actual operations performed by this code.
This problem has became evident lately with 
the widespread surreptitious deployment of cryptocurrency mining scripts in thousands 
of websites, exploiting the visitors' browsers without their
consent~\cite{js_mining, js_mining2}. Although there are some
blacklist-based extensions and tools that can protect users to some 
extent, such as Google's safe browsing~\cite{safebrowsing},
these do not offer complete protection.

On the other hand, disabling entirely the execution of \js code
often breaks intended legitimate functionality and affects the 
overall user experience. In general, the highly dynamic nature of \js,
the lack of mechanisms for informing users
about the implemented functionality, and the 
instant execution of script code, which does not leave room for extensive
security checks before invocation, 
are facilitators for malicious or unwanted in-browser code execution.

On the positive side, unwanted \js execution so far has been constrained
chronologically to the lifetime of the browser window or tab that rendered the 
compromised or malicious website. Consequently, cryptocurrency mining or other
malicious \js code can affect users only temporarily, typically for just
a few minutes~\cite{visitduration},
depending on the time a user spends on a given website.
Unfortunately, however,
some recently introduced web technologies---already supported by most 
popular browsers---can severely exacerbate the threat of unwanted \js computation in terms of stealthiness, persistence, and scale, and the support of such capabilities has already started raising concerns of the community~\cite{concerns1,concerns2}.

In this paper, we present \name: a system that enables a remote 
attacker to control users' browsers and hijack device resources.
Upon visiting a website that employs 
\name, the user's browser joins a centrally orchestrated
swarm that exploits user machines for unwanted computation,
and launching a wide variety of 
distributed network attacks. By leveraging the technologies offered by HTML5,
\name goes beyond existing approaches and demonstrates 
how malicious publishers can launch \emph{persistent} and \emph{stealthy}
attacks. This is possible by allowing malicious actors to continue
having control of the victim's browser \emph{even after 
the user browses away} from a malicious or infected website,
and by bypassing most of the existing in-browser detection mechanisms.

\name\ consists of two main parts: (a) an in-browser component, and 
(b) a remote command and control system. Although \name enables 
the attacker to perform attacks similar to those carried out by 
typical botnets~\cite{6616686}, there are some fundamental differences.
First and foremost, \name\ does not exploit \emph{any} implementation 
flaw on the victim's system and does not require the installation of \emph{any}
software. In contrast, \name, leverages the provided 
capabilities of \js and relies on some already available
HTML5 APIs. Consequently, \name\ is compatible with the vast majority 
of both desktop and mobile browsers. In contrast to previous approaches
for browser hijacking
(e.g., Puppetnets~\cite{puppetnets}), a key feature of \name\
is that it remains operational
even after the user browses away from the malicious web page.

In particular, our system fulfills three important objectives: (i) isolation 
from the visited website, allowing fine-grained control of the utilized
resources; (ii) persistence, by continuing its operation uninterruptedly on
the background even after closing the parent tab; 
and (iii) evasiveness, avoiding detection by browser extensions 
that try to monitor web page activity  
or outgoing communication. Besides malicious computation 
some of the attacks the 
infected browsers can perform include DDoS,
darknet creation, and malicious 
file hosting and sharing. 

In summary, in this paper, we make the following contributions:

\begin{enumerate}
	\item We present \emph{\name}: a novel multi-attack framework to allow
	persistent and stealthy bot operation through web browsers. \name
	is based on an in-browser execution environment that provides
	isolated execution, totally independent 
	from any open browsing session (i.e., browser tab). Therefore, it is able to withstand any tab crashes and shutdowns, significantly increasing the attackers firepower by more than an order of magnitude.
	\item We demonstrate and assess the feasibility of our approach by
	implementing a proof of concept prototype of \name for the 
	most common web browsers (i.e., Chrome, Firefox, Opera, and Safari). To measure its effectiveness, we thoroughly
	evaluate \name for various different attack scenarios.
	\item We discuss in detail various defense 
	mechanisms that can be applied as
	countermeasures against \name-like attacks.
\end{enumerate}

\noindent
The ultimate goal of this work is to raise awareness regarding the powerful
capabilities that modern browser APIs provide to attackers, so that a
more secure client-side environment can be provided for web applications in the future.

\section{Background}
\label{sec:background}

In this section, we discuss about several features that have been recently introduced as part of HTML5 and influence our design. We also discuss about the capabilities of web browsers' extensions, especially
with regards to these HTML5 features,
and finally, for each feature, we analyze its security aspects, access 
policies, permissions, and threat vectors that may open.

\subsection{HTML5 features}

\subsubsection{Web Workers}

Browsers typically have one thread that is shared for both the execution 
of \js\ and for page rendering processing. As a result, page updates 
are \emph{blocked} while the \js\ interpreter executes code, and vice 
versa. 
In such cases browsers typically ask the user whether to kill the 
unresponsive page or wait until the execution of such long-running 
scripts is over.  
HTML5 solves this limitation with the Web Workers
API~\cite{workerAPI},
which enables web applications to spawn background workers for 
executing processing-intensive code in separate threads from the 
browser window's UI thread. 

Since web workers run as separate threads, isolated from the 
page's window, they do not have access to the Dynamic Object Model 
(DOM) of the web page, global variables, and the parent object variables 
and functions. More specifically, neither the web worker can access its 
parent object, nor the parent object can access the web worker. Instead, 
web workers communicate with each other and with their parent object 
via message passing.
Web workers continue to listen for messages until the parent object terminates 
them, or until the user navigates away from the main web page.
Furthermore, there are two types of web workers: dedicated and shared 
workers. Dedicated web workers are alive as long as the parent web page 
is alive,
while shared web workers can communicate with 
multiple web pages, and they cease to exist only when 
all the connections to these web pages are closed. 

Typically, web workers are suitable for tasks
that require computationally intensive processing
in an asynchronous and parallel fashion, 
such as parsing large volumes of data and performing computations 
on arrays, processing images and video, data compression,
encryption etc. Indeed, during the recent outbreak of web-based
cryptocurrency mining, we have observed that typically these scripts 
utilize web workers for mining, and that they deploy multiple such 
workers to utilize all available CPU cores of the user's system.  

\subsubsection{Service Workers}
\label{sec:sw}

Service workers are non-blocking (i.e., fully asynchronous) modules
that reside in the user's browser, in between of the web page and the web server.
Unlike web workers, a service worker, once registered and activated, 
can live and run in the background, \emph{without} requiring 
the user to continue browsing through the publisher's website---service
workers run in a separate thread and their lifecycle is completely 
independent from the parent page's lifecycle.
%
The characteristics of service workers enable the provision of functionality
that cannot be implemented using web workers,
such as push notifications and background syncing with the publisher.
Furthermore, another core feature of service workers 
is their ability to intercept and handle network requests, including 
programmatically managing the caching of responses. This allows service 
workers to be used as programmable network proxies, allowing 
developers to enrich the offline user experience by controlling how network
requests from a web page are handled.

A service worker can be registered only over HTTPS via the
\texttt{serviceWorkerContainer.register()} function, which
takes as argument the URL of the remote \js\ file that contains the worker's script. 
This URL is passed to the internal browser's engine and is fetched from there.
For security purposes, this \js file can be fetched only from the first-party domain (i.e., it cannot be hosted in a CDN or any other third-party server). Also, no iframe or third-party script can register its own service worker.
Furthermore, importantly, no browser extension or any in-browser entity can have access either in the browser's C++ implementation that handles the retrieval and registration of the service worker or in the first-party domain.

When the user browses away from a website, the service worker of that website is typically paused by the browser; it is then restarted and reactivated once the parent domain is visited again. However, it is possible for the publisher of a website 
to keep its service worker alive
by implementing periodic synchronization.
It should be noted though that the registration of a service worker is entirely non transparent to the user, as the website \emph{does not} require the user's permission to register and maintain a service worker. Furthermore, similarly to web workers, service workers cannot access the DOM directly. Instead, they communicate with their parent web pages by responding to messages sent via the \texttt{postMessage} interface.

\subsubsection{WebRTC}
\label{sec:webrtc}

Popular web-based communication applications (e.g., Web Skype, Google Meet, Google Hangouts, Amazon Chime, Facebook Messenger) nowadays are based on
Web Real-Time Communication (WebRTC) API~\cite{webrtc2}, which enables
the establishment of peer-to-peer connections between browsers. With WebRTC, browsers can
perform real-time audio and video communication and exchange data
between peers, without the need of any intermediary.

As in every peer-to-peer protocol, a challenge of WebRTC is to locate and
establish bidirectional network connections with remote peers
residing behind NAT.
To address this, WebRTC uses STUN (Session Traversal Utilities for NAT) and
TURN (Traversal Using Relays around NAT) servers for resolving the network
address of the remote peer and reliably establishing a connection.
There are several such servers publicly available~\cite{STUNfree},
maintained either by organizations (e.g., Universities) or companies (e.g.,
Google).

\subsubsection{Cross-Origin Resource Sharing}
\label{sec:cors}

Before HTML5, sending AJAX requests to external domains was impossible due to
the restrictions of the same-origin policy, which restricts scripts running as part 
of a page in accessing only the DOM and resources of the same domain.
This means that a web application using AJAX APIs (i.e., \texttt{XMLHttpRequest}
and the Fetch API) can only request resources from the same domain it
was loaded.

However, the Cross-Origin Resource Sharing (CORS)~\cite{cors} capabilities introduced in HTML5, allow scripts to make cross-origin AJAX requests to other
domains.
To enable this feature, CORS uses extra HTTP headers to permit a user agent
to access selected resources from a server on a different domain (origin) than
the parent sits.
Additionally, for HTTP request methods that can cause side-effects on
server-side data (in particular, for HTTP methods other than \texttt{GET}, or
for \texttt{POST} usage with certain MIME types), the specification mandates
browsers to ``preflight'' the request, soliciting supported methods from the
server with an HTTP \texttt{OPTIONS} request method, to determine
whether the actual request is safe to send~\cite{cors}.

\subsection{Web Extensions}
\label{subsec:ext}

The current design of modern browsers' extensions allows 
	two types of \js scripts within a browser extension: (a) content 
	scripts and (b) background scripts. Content 
	scripts~\cite{contentscripts} run in the context of the 
	websites visited by the user, thus they can read and 
	modify the content of these websites using the 
	standard DOM APIs~\cite{domapis}, similarly to the 
	websites' scripts (i.e., \js scripts included by 
	the publisher). Furthermore, content scripts can 
	access directly a small subset of the 
	WebExtension \js APIs~\cite{webextapis}.

On the other hand, background scripts run as long as the 
	browser is open (and the extension is enabled), and typically 
	implement functionalities independent from the lifetime of any 
	particular website or browser window, and maintain a 
	long-term state. These background scripts cannot access 
	directly the content of the websites visited by the user. 
	However, background scripts can access all the 
	WebExtension \js APIs (or chrome.* APIs for Google 
	Chrome), if the user's permission is granted during the 
	installation of the extension.

Indicatively, the large set of WebExtension \js APIs 
	contains the bookmarks, cookies, history and storage 
	APIs, which allow access on various types of user data, 
	the tabs and windows APIs, browserSettings and the 
	webRequest API among many others. A list of the 
	available WebExtension APIs and information regarding 
	their support by major browsers can be found in \cite{jsapis}. 
	However, even though content scripts cannot access 
	all WebExtension APIs directly, and background 
	scripts cannot access the content of the visited 
	website, this can be achieved indirectly since 
	the content and background scripts of an 
	extension can communicate with each other. 

In addition to the above mentioned APIs, Google Chrome 
	also supports some HTML5 and other emerging APIs for its extensions, 
	such as audio, application cache, canvas, 
	geolocation, local storage, notifications, video and web 
	database~\cite{chromeext}. However, it is important with 
	regards to this work to emphasize that \emph{none of the 
		browsers allow extensions to use HTML5 APIs} such 
	as the Service Workers or the Push API. 
	Consequently, browser extensions cannot interact with 
	possible deployed service workers in any way, (e.g., modify their code, monitor their outgoing traffic, etc.).

\begin{table*}[t]
	\caption{Analysis of HTML5 \js\ execution methods.}
	\label{tab:html5_analysis}
	\centering
	\begin{tabular}{lccccccc}
		\toprule
		\multicolumn{1}{c}{\textbf{Feature}} &
		\begin{tabular}[x]{@{}c@{}}\textbf{Concurrent}\\\textbf{Execution}\end{tabular} & 
		\begin{tabular}[x]{@{}c@{}}\textbf{Background}\\\textbf{Execution}\end{tabular} & 
		\begin{tabular}[x]{@{}c@{}}\textbf{Webpage}\\\textbf{Detached}\end{tabular} & 
		\begin{tabular}[x]{@{}c@{}}\textbf{Intercept}\\\textbf{HTTP Requests}\end{tabular} & 
		\begin{tabular}[x]{@{}c@{}}\textbf{Persistent}\\\textbf{Storage}\end{tabular} & 
		\begin{tabular}[x]{@{}c@{}}\textbf{DOM}\\\textbf{Access}\end{tabular} & 
		\begin{tabular}[x]{@{}c@{}}\textbf{Network}\\\textbf{Access}\end{tabular}\\ 
		\midrule
		\rowcolor{Gray}
		\textbf{Local \js\ code}       &               &              &                 &    \cmark  &  \cmark  &  \cmark &  \cmark \\
		\textbf{Web Worker (Shared)}         &   \cmark  &              &                 &                &  \cmark  &             &  \cmark \\
		\rowcolor{Gray}
		\textbf{Web Worker (Dedicated)}      &   \cmark  &              &                 &                &  \cmark  &             &  \cmark \\
		\textbf{Service Worker}              &   \cmark  &  \cmark  &    \cmark   &    \cmark  &  \cmark  &             &  \cmark \\ \bottomrule
		
	\end{tabular}%
\end{table*}

\subsection{Security Analysis}

Table~\ref{tab:html5_analysis} summarizes the characteristics
of different APIs of interest.
We categorize them along four axes related to the efficiency of a distributed botnet:
(i) the execution model (i.e., whether it can run in parallel to the main web page or in the background),
(ii) if direct network access is possible,
(iii) the ability to use persistent storage,
and (iv) the ability to access the DOM of the web page.

\js\ code (running either as part of a web page, or in a
web worker or service worker) has access to persistent storage
(e.g., using WebStorage API~\cite{webstorage}),
as well as the ability to communicate with other servers or peers (e.g.,
using the XHR~\cite{xhr}, WebSockets~\cite{websocket}, or WebRTC~\cite{webrtc} APIs).
However, local \js\ code embedded in the webpage also has direct access to
the page's DOM and therefore, the ability to access or manipulate any element of the webpage, as well as
any network request or response that is sent or received.
Page-resident \js\ code cannot be detached from the webpage, neither run without
blocking the rendering process.
This results to a major limitation (for the purposes of malicious scripts),
as long-running operations would affect the user experience.
Also, a suspicious code snippet could be detected easily by 
browser extensions, since it needs to be embedded in the main website, and extensions' \js code can access, inspect and 
in general, interfere with the content of the visited website.

Web workers, on the other hand, can perform resource-intensive operations without affecting the user's browsing experience, as they run in separate threads. This allows utilizing all different available CPU cores of the user's machine, by spawning a sufficient number of web workers.
Service workers behave in a similar fashion, but have the important advantage of being completely detached from the main web page, running in the background even after the user has navigated away.
Moreover, service workers can intercept the HTTP requests sent by 
the web page to the back-end web server. Importantly, since service workers are completely detached from the page's window, extensions cannot monitor or interfere with them.

Finally, using the CORS capabilities of HTML5, it is possible to send multiple GET or POST requests
to third-party websites.
However, the \texttt{Access-Control-Allow-Origin:*} header has to be set
by the server, in order for the request to be able to fetch any content.
Besides sending HTTP requests, the WebRTC API~\cite{webrtc} allows the
peer-to-peer transfer of
arbitrary data, audio, or video---or any combination thereof.
This feature can open the window for malicious actions such as illegal hosting
and delivery of files, as well as anonymous communication through a network
of compromised browsers, as we showcase later on.

\section{Threat Model and Objectives}
\label{sec:objectives}

The motivation behind this work is to design a system capable of turning
users' browsers into a multi-purpose ``marionette'' controlled by a malicious remote entity. 
Our goal is to leverage \emph{solely} existing HTML5 features
in order to highlight the lack of adequate security controls in 
modern browsers that would have prevented the abuse of these advanced features.

\subsection{Threat Model}
\label{sec:threatmodel}

We assume a website that delivers malicious content
to execute unwanted or malicious
\emph{background} operations in visitors' browsers. Once the website is rendered, this
malicious content is loaded in a service worker that is capable of continuing its operation
even \emph{after the victim browses away from the website}.

Websites can deliver such malicious or unwanted content \emph{intentionally},
to gain profit directly (e.g., by attracting visitors and thus
advertisers),
or indirectly, by infecting as many user browsers as possible
to carry out distributed (malicious) computations or mount large-scale attacks.
There are also several cases where a website can end up 
hosting malicious content \emph{unintentionally}. Those cases include:
(i) the website registers a benign service worker that includes untrusted
dynamic third-party scripts~\cite{Lekies:2015:UDD:2831143.2831189}, which in turn possibly load malicious code;
(ii) the website includes third-party libraries\footnote{Modern websites often include numerous third-party scripts~\cite{thirdParties1,thirdParties2} for analytics or user tracking purposes, aiming to gain insight, improve performance, or collect user data for targeted 
advertising.}, one of which can turn rogue or be compromised, and then divert
the user to a new tab (e.g., using popunders~\cite{popunder2}
or clickjacking~\cite{clickjacking}) where it can 
register its own service worker bound to a \emph{third-party} domain;
(iii) the website is compromised and attackers plant their malicious \js\ code directly
into the page, thus registering their malicious service worker 
--- a scenario that we see quite often in recent years~\cite{showtime,ronaldo}; or
(iv) the website includes iframes with dynamic content, which are typically auctioned
at real-time~\cite{auctions} and loaded with content from third parties.

In the latter case,
malicious actors can use a variety of methods (e.g.,
redirect scripts~\cite{breakout,6956553} or social
engineering) to break out of the iframe and open a new tab on the user's browser
for registering their own service worker.
The important advantage of this latter approach is that the user does not need to re-visit the website 
for the service worker to be activated. After registration, just an iframe
loaded from the malicious third party is enough to 
trigger the malicious service worker, regardless of the visited first-party website.
This relieves the attackers from the burden of maintaining websites with
content attractive enough to lure a large number of visitors. Instead, attackers can activate their bots just by running malvertising campaigns, purchasing iframes in ad-auctions~\cite{malvertising}.



	
\begin{figure*}
	\centering
	\includegraphics[width=1.8\columnwidth]{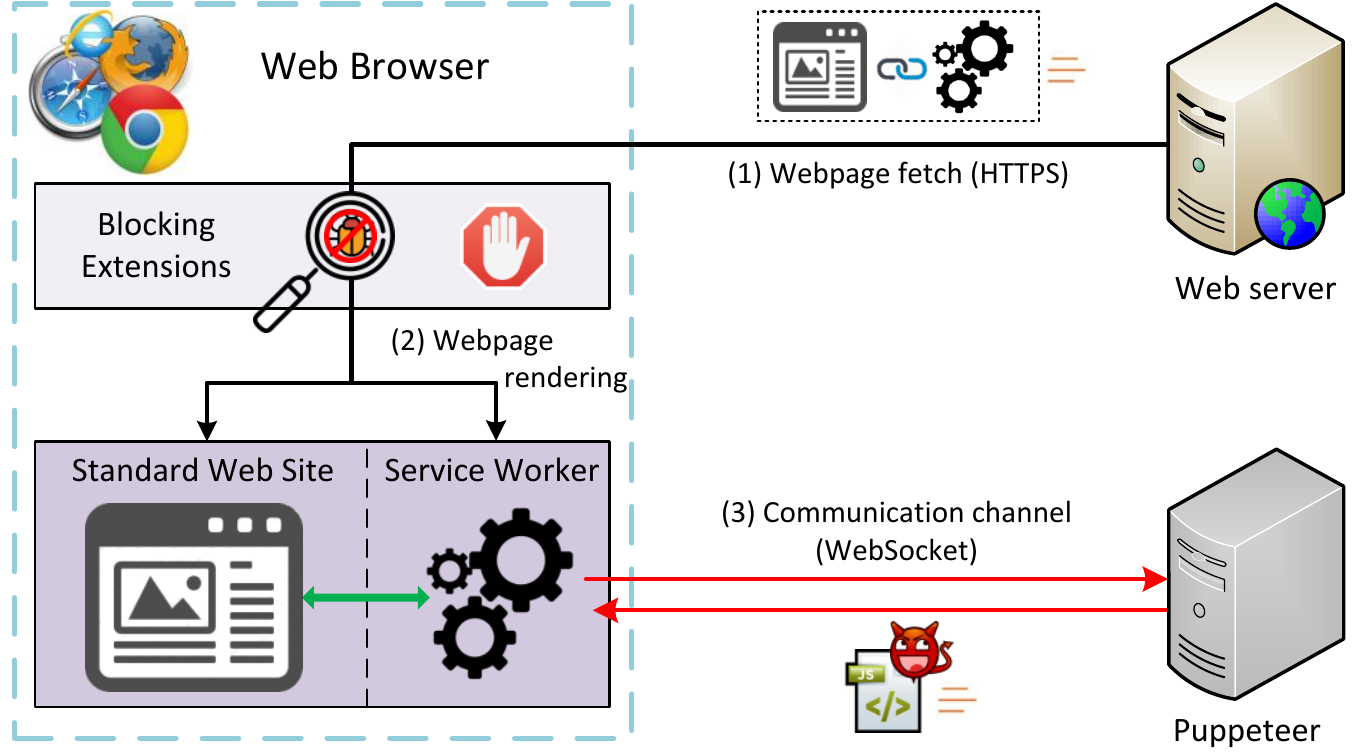}
	\caption{High level overview of \name. The in-browser component (Servant), embedded in a Service Worker, gets delivered together with the actual content of a website. After its registration on the user's browser, it establishes a communication channel with its remote command and control server (Pupeteer) to receive tasks.}
	\label{fig:overview}
\end{figure*}

\subsection{Challenges}
The greatest challenge for systems like \name is to keep the user's device under
control
for as long as possible. This is a challenging task given that a connection
with the server may be possible only for the duration of a website visit;
recent studies have estimated the average duration of a typical website visit
to be less than one minute~\cite{visitduration}.
In addition, there is a plethora of sophisticated browser extensions~\cite{tamperData,mcafeeSecure,privacyManager,adblockplus,ghostery,purevpn} that 
monitor the incoming and outgoing traffic of in-browser components.
Consequently, 
another challenge for \name is to evade any such deployed
countermeasure installed in the browser.
Finally, it is apparent that the malicious or unwanted computation of \name must not impede
the normal execution of the browser, and avoid degrading the user experience.
Otherwise, the risk of being detected by a vigilant user or at least raising suspicion
due to reduced performance is high.

To summarize, in order to overcome the above challenges, a \name-like system should have the following properties:

\begin{enumerate}
\item \emph{Isolation:} the system's operation must be independent from a browsing
session's thread or process. This isolation will allow a malicious actor to
perform more heavyweight computation without affecting the main
functionality of the browser.

\item \emph{Persistence:} the operation must be completely detached from any ephemeral browsing session, so that the browser can remain under
the attacker's control for a period longer than a short website visit.

\item \emph{Evasiveness:} operations must be performed in a stealthy
way in order to remain undetected and keep the browser infected as long as
possible.
\end{enumerate}

\section{System Overview}
\label{sec:design}

In this section, we describe the design and implementation of
\emph{\name}, a multi-purpose web browser abuse infrastructure, and present in detail how we address the
challenges outlined earlier.
Upon installation, \name allows a malicious actor
to abuse computational power from users' systems through their browsers, and perform a variety of unwanted or malicious activities.
By maintaining an open connection with the infected browser, the malicious actor can change the abuse model at any time, instructing for instance an unsuspecting user's browser to switch from illicit file hosting to distributed web-based cryptocurrency mining.

Our system, which is OS agnostic, assumes \emph{no} assistance from the user
(e.g., there is no need to install any browser extension). 
On the contrary, it assumes a ``hostile'' environment with possibly more than
one deployed anti-malware browser extensions and anti-mining countermeasures.
We also assume that \name\ targets off-the-shelf web browsers. Hence, the execution
environment of \name\ is the \js\ engine of the user's web browser. Breaking
out of the JIT engine~\cite{JSbreak} is beyond the scope of this paper.

\subsection{System components}
Figure~\ref{fig:overview} presents an overview of \name, which consists of
three main components:

\begin{enumerate}
  \item {\bf Distributor:} a website under the attacker's control (e.g.,
    through the means discussed in Section~\ref{sec:threatmodel}), which
    delivers to users the \name's Servant component, along with the
    regular content of the web page.
	It should be noted that the attacker does not need to worry about the time a user will spend on the website. It takes only one
	visit to invoke \name\ and run on the background as long as the victim's browser is open.

	\item {\bf Servant:} the in-browser component of \name, embedded in
	  a service worker. It gets delivered and planted inside the user's web browser
	  by the Distributor.
	Upon deployment, the Servant establishes a connection with its Puppeteer
	through which it sends heartbeats and receives the script of the malicious
	tasks it will perform.
	The Servant runs in a separate process and thereby it continues its operation uninterruptedly even after its parent browsing tab closes. 
	
	\item {\bf Puppeteer:} the remote command and control component. This component sends
	  tasks to the Servant to be executed, and orchestrates the performed malicious operations.
	  The Puppeteer is responsible for controlling the intensity of resources utilization (CPU, memory, etc.) on the user side, by tuning the
	computation rate of the planted Servant. 
\end{enumerate}

As illustrated in Figure~\ref{fig:overview}, \name\ is deployed in three main steps: 
First, (step~1) the user visits the website (i.e., the Distributor)
to get content that they are interested in.
The Distributor delivers the \js\ code of the Servant along with the
rest of the webpage's resources.
During webpage rendering (step~2), the Servant is deployed
in the user's browser. As part of its initialization,
the Servant establishes a communication channel with its remote command and
control server (Pupeteer) and requests the initial set of tasks (step~3). 
The Pupeteer, maintained by the attacker, responds with the malicious script
(e.g., DDoS, password cracking, cryptocurrency mining) the Servant has to execute. 

\subsection{Detailed Design}
\name leverages existing components of HTML5 to achieve the objectives
presented in Section~\ref{sec:objectives}: isolation, persistence, and evasiveness.
%
In-browser attacks that involve computationally heavy workloads require
\emph{isolation} in order to avoid interfering with a webpage's core
functionality. Previous approaches~\cite{dorsey,pan2016assessing}
rely on web workers~\cite{workerAPI} to carry out heavy computation
in the background (in a separate thread from the user's interface scripts).
Although this isolation also prevents the code of the web worker from having
access to the DOM of the parent page, it has the benefit of allowing multi-core utilization.
As a result, attackers can utilize simultaneously many cores for their malicious computations.
However, web workers run in the same browser tab as the website, and
consequently, their execution is tightly coupled with the parent tab: whenever the tab closes, 
the web worker terminates as well. In addition, security-related browser extensions
can (i) monitor all traffic and (ii) tamper with the script running in the 
web worker.

\begin{figure*}[t]
	\begin{minipage}[t]{0.33\textwidth}
		\centering
		\subfigure[]{
			\label{fig:usecase1}
			\includegraphics[width=0.95\hsize]{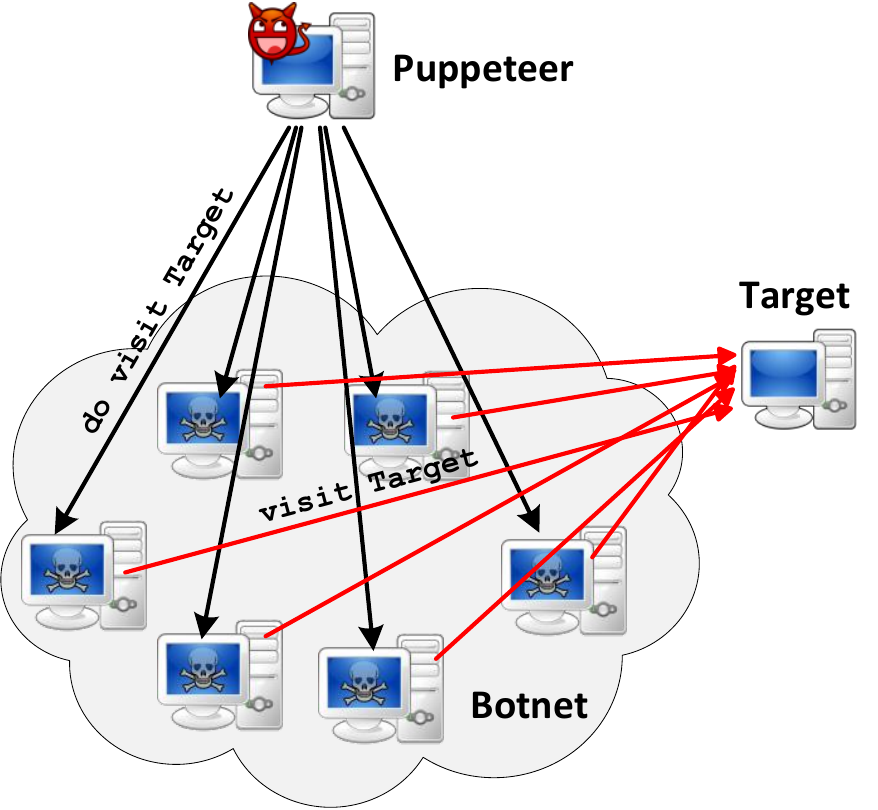}
		} 
	\end{minipage}
	\begin{minipage}[t]{0.33\textwidth}
		\centering
		\subfigure[]{
			\label{fig:usecase2}
			\includegraphics[width=0.95\hsize]{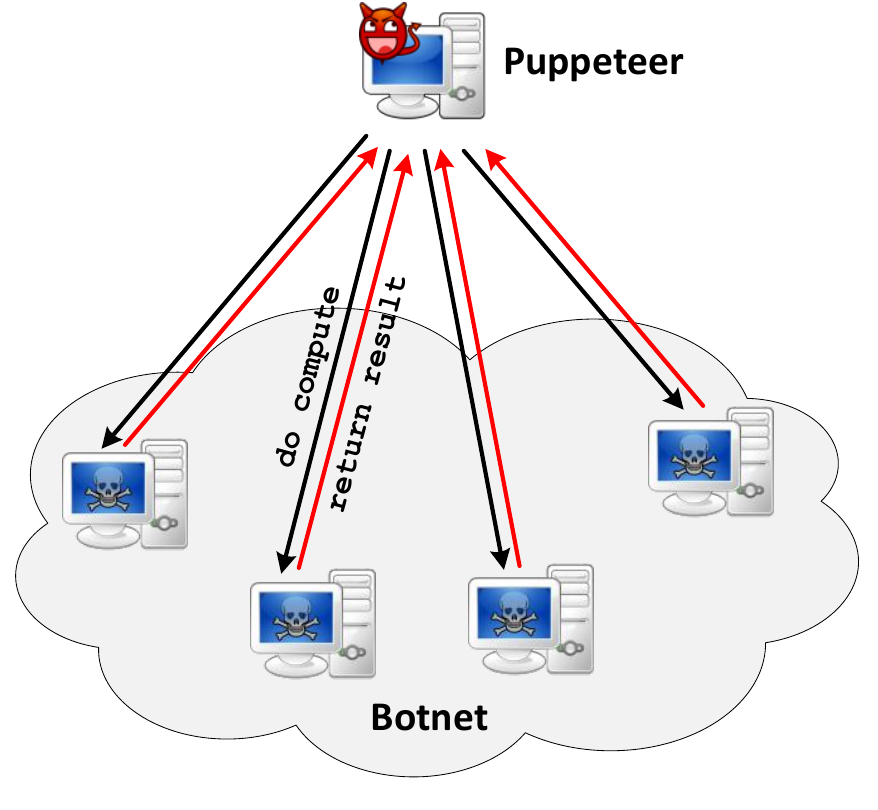}
		}
	\end{minipage}
	\begin{minipage}[t]{0.33\textwidth}
		\centering
		\subfigure[]{
			\label{fig:usecase3}
			\includegraphics[width=0.95\hsize]{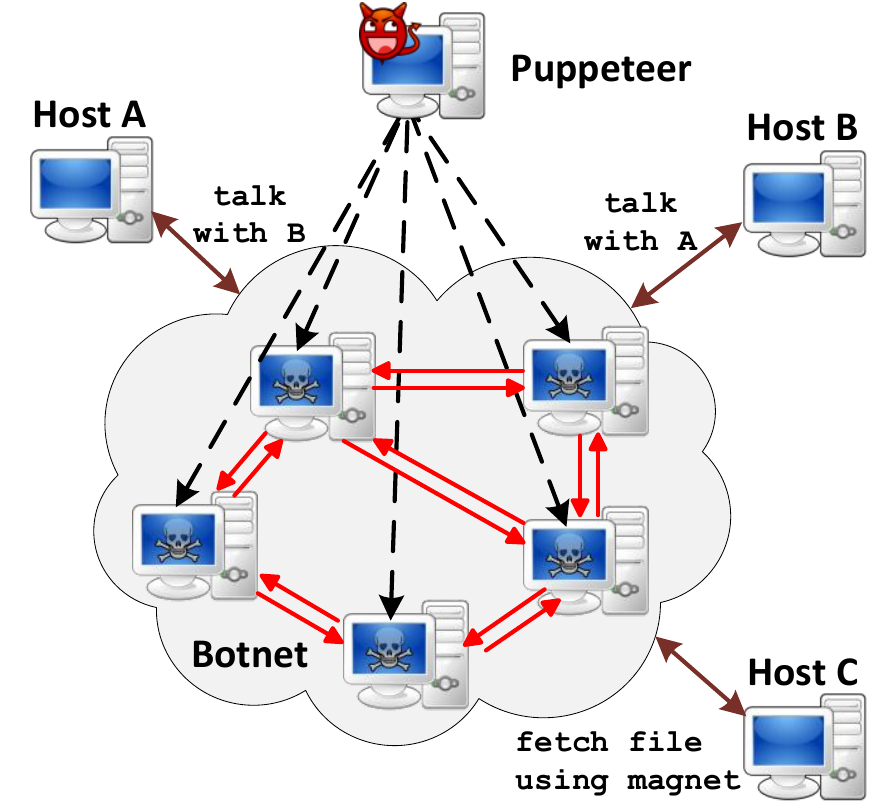}
		}
	\end{minipage}
	\caption{Different use cases of \name. After victims get compromised, the attacker can
		instrument them to perform (a) visits to a selected server or URL, for DDoS attack or fake ad-impressions, (b) requested computations, such as
		cryptocurrency mining or password cracking, and (c) illegal services, such as illicit file hosting or hidden/anonymized communications.}
	\label{fig:usecases}
\end{figure*}

To remedy these shortcomings, \name leverages a different, relatively recent
feature of HTML5, namely \emph{service workers}~\cite{serviceWorker}.
As described in Section~\ref{sec:sw}, service workers are typically used as an in-browser caching proxy,
serving the user during offline periods. In contrast to web workers, service
workers run in a separate process, completely detached from the parent tab.
In addition to service workers, we use the SyncManager
interface~\cite{syncManagerAPI} to register background ``sync registrations''
for the service worker, to keep the Servant always alive.
The tab independence and indefinite lifetime properties of the Servant provide
\name with \emph{persistence}, allowing attackers to carry out their malicious
computation for the entire period that a browser remains open---a major
benefit over existing approaches based on web workers,
which remain operational only for the duration of a browsing session (open tab).

Another advantage of leveraging service workers is that they conceptually
operate between the browser and the remote server. As a consequence,
any security monitoring performed by additional browser extensions cannot
monitor the activity and network communication of the service worker,
allowing the Servant to operate in a stealthy way.
Consequently, the Servant
can establish a communication channel with the remote Puppeteer that no browser extension
can snoop. In addition, the established communication channel is TLS-encrypted (as required by the service worker API~\cite{sslSW}). This way, the integrity and the confidentiality of the transmitted data is ensured.
Consequently the C\&C communication channel cannot be inspected by any eavesdropping 
third party sitting either (i) inside (e.g., browser extension) or (ii) outside (e.g., ISP) of the browser.

The only request that reveals the existence of the service worker is the initial GET request 
at the time of the user's first website visit, when the service worker gets initially registered.
Although during that GET request a monitoring extension can observe the contents of the 
service worker, it will still not observe any suspicious code---the code that
will carry out the malicious tasks is delivered to the Servant
only after its first communication with the Puppeteer, 
and this communication is hidden from browser 
extensions (as discussed in Section~\ref{subsec:ext})

Along with the evasiveness of \name\ against monitoring and blocking extensions, it is also important to maintain its stealthiness to avoid detection from users themselves.
Existing web-based botnet approaches~\cite{dorsey,pan2016assessing} follow an opportunistic 
approach, utilizing greedily all available resources on the device during
their limited period of activity.
When browsers run such a malicious script,
the louder noise of the fans, the sudden power drainage, or the
sluggish responsiveness of their system, alerts users who are likely to close the associated browser tab, or even report the website to their blocker extension.

In contrast to existing in-browser attacks,
\name aims to prolong its presence on a user's device by
allowing the attacker to monitor the state of the device, and adjust
accordingly the resources utilization in order not to raise any
suspicion. To that end, the periodic heartbeats sent from Servant to Puppeteer also contain (i) the
current CPU utilization of the device using HTML5's hardwareConcurrency API~\cite{hardConcurrency} and (ii)
its current battery status (``charging'' or ``on battery'' and battery level)
using the Battery Status API~\cite{battery}.
Having this information, the Puppeteer can reduce or pause the malicious
workload, and minimize the risk of self-exposure in case of a possible CPU capping mechanism of the browser~\cite{chrome-throttling,firefox-throttling}.

\subsubsection*{Persistence across Browser Reboots}
\name\ runs in the background as long as the browser is open.
After that, the victim has to re-visit the malicious domain or render the malicious iframe where the malicious domain resides,
in order to re-activate the service worker and allow the Servant to continue
its operation.

To increase persistence even further, we have developed a technique that
allows \name\ to persist even after the browser has been restarted. This can be achieved by utilizing the Push API~\cite{push}. 
This feature allows a web server to deliver asynchronous notifications and updates to service workers, in an attempt to provide users with 
better engagement and timely new content. By abusing this mechanism, \name\ can enable the Puppeteer to periodically probe its Servants and re-activate 
them after the browser restarts.

In contrast to the non-transparent to user process of service worker registration, security policies in modern browsers restrict the use 
of the Push notifications feature only after the user's permission.
Of course, some users may get suspicious on that behavior, depending on the website they visit. However, an advanced attacker can convince reluctant 
users to give their consent for push notifications by advertising enticing
offers (e.g., virtual points or participation to contests) or by performing
more advanced types of social 
engineering using custom permission requesting popups. Recent studies have
shown that 12\% of users give such permissions when they are asked 
to~\cite{usersPush}, which constitutes a fairly large number of nodes, sufficient for deploying a persistent botnet that is capable to survive browser reboots.

\section{Attack Vectors}
\label{sec:attacks}

Our design, described in Section~\ref{sec:design}, opens the space
for a diverse set of attacks in users' web browsers, which can be
categorized in three models, as shown in Figure~\ref{fig:usecases}.

\subsection{DDoS Attacks}
\label{sec:ddos}
A simple yet powerful attack that can be launched with the devices the attacker controls is a Distributed Denial-of-Service attack. In \name\ we implemented a DDoS attack module enabling the Puppeteer to instruct the Servants to connect to a specific Internet host. As a result, the targeted host will get overwhelmed by the large amount of connections and thus become non-responsive to benign users.

A limitation of using a high-level language, such as \js, to initiate
a DoS attack is that it does not provide low level networking access.
Directly manipulating the network packets
to be sent is thus not an option (e.g., force
TCP-SYN packets only, or spoof source network address).
In addition, it results to much higher latency, due to the extra
memory copies and context switches that are caused from the resulting system calls.
Instead, \js\ offers more high-level approaches, such as
\texttt{XMLHttpRequest} objects~\cite{xhr} or methods provided by cross-platform libraries (e.g., the
\texttt{get()}, \texttt{post()}, and \texttt{ajax()} methods provided by jQuery).
These methods can be used to perform HTTP GET and POST requests, either
synchronously (i.e., in a blocking
fashion, waiting for the connection to be established)
or asynchronously.
In addition, some methods may return cached responses (e.g., the \texttt{get()}
method provided by jQuery).

In order to increase the DDoS fire power of \name, we
use the \texttt{XMLHttpRequest} API,
which can be used to perform AJAX (asynchronous HTTP) requests, and does not
cache any responses.
Moreover, it allows to control the
request method, and set an arbitrary HTTP body, as well as some HTTP request
headers (e.g., the request content type).
One concern though, that we already mentioned in Section~\ref{sec:cors}, is that
if the
target web server does not enable the \texttt{Access-Control-Allow-Origin:*}
header, the request will not fetch any content.
Even in that case though, the attack can still succeed, as it does not
necessarily rely on forcing the web server to send a response.
As long as the requests are sent, the incoming network link is filling up
and also the server needs to spend resources to handle the incoming requests.

Apart from HTTP fetching mechanisms, HTML5's WebSockets API~\cite{websocket}
provide additional opportunities.
WebSockets can be used to send messages to a WebSocket-enabled server
over TCP and receive event-driven responses.
Obviously, to mount a DoS attack using WebSockets, the targeted server
needs to implement this protocol; this is indeed the case for many
popular web sites, as well as for smaller ones, which increasingly adopt
the WebSockets protocol.
Besides that, as already has been shown in~\cite{191940}, malicious \js\
code may still misuse the handshake by requesting resources even by targeting
a non-WebSocket web server.
Although the targeted web server may ignore the characteristic WebSocket HTTP
headers (as it is not supported), it can still accept WebSocket handshake
HTTP requests as normal HTTP requests~\cite{191940}.
As a result, the web browser will start the WebSocket handshake with the
target, while the non-WebSocket web server will process the HTTP request
as a valid request.
In \name, we use the \texttt{WebSocket()} method to initiate 
connections with web servers, and then the \texttt{send()} method to send
a flood of data to the targeted server.

Using \texttt{XMLHttpRequest.send()}, jQuery's \texttt{ajax()} and WebSocket's \texttt{send()}
methods, we can continuously send a flood of messages to a targeted host.
Each approach allows \name to connect to any host, by
specifying the hostname or IP address and the corresponding
port number. By doing so, \js\ code can misuse the TCP handshake by requesting
connections even to non-HTTP or non-WebSocket servers.
In those cases, the targeted servers will either receive only the TCP SYN
packets (e.g., when the destination port is in a closed, reject, or drop state),
or the full HTTP request.
Furthermore, the WebSocket API allows to open many different connections, which enables attackers to orchestrate
different styles of attacks (e.g., stealthy, low-volume, etc.).
For instance, it allows to perform Slowloris-like attacks, by keeping many connections to the target server open as long as possible~\cite{slowloris}.

Of course, \name\ cannot send messages to any port at the targeted host.
To avoid Cross-protocol Scripting~\cite{cps}, which allowed the transmission
of arbitrary data to any TCP port,
modern browsers block by default outgoing messages to a list of reserved ports~\cite{closedports}.
Finally, we note that the resulting network performance of \js\
is not that high, compared to DoS attack tools that can leverage
direct access to OS internals (i.e., memory map techniques between the 
network interface and the application) and low-level APIs (i.e., raw sockets).
However, this is not a serious limitation, as it has been shown that short,
low-volume DDoS attacks pose a great security and
availability threat to businesses~\cite{lowvolumedos}.

\subsection{Cryptocurrency Mining}

The rise of lightweight cryptocurrencies, such as JSEcoin and
Monero, together with the features of Web Workers API that have
been described in Section~\ref{sec:background}, have recently
enabled the widespread adoption of
cryptocurrency mining on the Web.
As a result, attackers have started migrating mining algorithms
to \js\ and embed them to regular websites, in the form
of web worker tasks.
By doing so, the website visitors become mining bots unwittingly
every time they access these websites.

However, the short website visiting times make the profitability
of the web workers approach questionable~\cite{webminingwaste}.
Instead, \name increases the potential profits
of web cryptocurrency mining, due to the background execution it offers,
completely detached from the website.
As a matter of fact, we have implemented a service worker module
that computes hashes of the popular CryptoNight algorithm~\cite{cryptonight}.
CryptoNight is a proof-of-work (PoW) algorithm used in several cryptocurrencies,
such as Electroneum (ETN) and Monero (XMR).
The service worker that we have implemented within \name, connects
with Coinhive~\cite{coinhive}, which is a web
service that provides an API for users to embed a \js\ miner
on their websites.
Alternatively, the cryptocurrency miner can connect to any mining
pools, through the HTTP stratum proxy, using a registered account,
as shown in previous works~\cite{pan2016assessing}.
By doing so, attackers will be credited the payout directly to
their wallets.
Finally, we notice that other hash algorithms used in for cryptocurrency mining, such as Scrypt-based miners~\cite{scrypt}, can be implemented in a straightforward way, by porting their implementations to \js.

\subsection{Distributed Password Cracking}
The idea of distributed password hash cracking on the web is not new~\cite{puppetnets}.
Orthogonal to other approaches that try to boost the sustained performance
by either increasing the parallelism using different web
workers~\cite{dorsey}, or exploiting the computational capabilities of modern
GPUs using the WebGL/WebCL API~\cite{webCLcracking},
\name can help towards increasing the uptime of hash cracking techniques, and as a result the overall performance.

The basic concept in \name is to have the Puppeteer distribute the computation
between the infected browsers.
The server contains a list of the hashes to be cracked and gives each node
a range of character combinations along with the hash to be cracked.
Each node then hashes these combinations and checks if it matches
the original hash; if it matches, the node reports the recovered password back to
the Puppeteer.
A major advantage of \name is that it can be agnostic to the hashing function
used, since the function code is transferred from the Puppeteer and executed
from the \name nodes through \texttt{eval()}.
As a matter of fact, in Figure~\ref{fig:password_hashing} we show the 
performance achieved by \name for executing two popular hashing algorithms, namely SHA-256 and MD5.

\subsection{Malicious or Illegal Data Hosting}
Having a large network of \name nodes can also enable the delivery of illegal or
otherwise unwelcome content.
The advantages of \name is not only that the content can be served by
unsuspecting users, making it hard to track down the real culprits behind it,
but also allows efficient data distribution between the \name nodes.

Indeed, the release of WebRTC (Web Real-Time Communications) protocol in the
browser a few years ago, enables peer-to-peer networking communications.
In particular, WebRTC allows Web applications and sites to capture and optionally
stream audio and/or video media, as well as to exchange arbitrary data between
browsers without requiring an intermediary.
Even though this technology opens new opportunities for distributed networking
to the web, it also brings some significant security concerns when used
maliciously.
In the case of \name, for instance, it could be easily used as an illegal
content provider, leveraging the distributed nature and persistence that
offers.
As a proof-of-concept, similar to~\cite{dorsey} we used the WebTorrent API~\cite{webTorrentAPI}
to implement
a simple, yet flexible, data hosting mechanism over WebRTC which allows the sharing of
torrent files through the infected \name nodes.
WebTorrent allows users to seed and leech files with other peers entirely
through their web browsers.
A new torrent file can easily be created using the \texttt{seed()} function
which creates a new torrent and starts seeding it. The file can then be
downloaded and further seeded from other nodes, using the returned \texttt{magnetURI}.

\subsection{Other Attacks}
\subsubsection{Relay Proxies}
Fully anonymous and transparent relay proxies that can route
data between two peers, are an important asset for criminal use, making it
difficult for the authorities to track down the perpetrators.
Large groups of such proxies can 
form a hidden network (i.e., Darknet~\cite{darknet}), where people 
buy and sell illicit products like weapons and drugs~\cite{DBLP:journals/corr/abs-1712-10068,martin2014drugs}.

The \name\ infrastructure can provide a platform for establishing such networks.
Specifically, an infected browser can be used as an intermediate proxy to fetch illegal content 
from services in the Darknet on behalf of an anonymous user. 
Indeed, building upon the previous illegal data hosting scenario, \name\ could
form anonymous circuits (similar to mixnets), through which users could route their web traffic. 
Such chain could be created by bots connected with encrypted peer-to-peer channels with each other by using WebRTC\footnote{WebRTC traffic is always encrypted. 
Transmitted data are protected by Datagram Transport Layer Security (DTLS)~\cite{dtls} and Secure Real-time Transport Protocol (SRTP)~\cite{srtp}.}.
There are already such browser-based proxies implemented over WebRTC, like Stanford's Flash Proxies~\cite{flashproxy} and Tor Project's Snowflake~\cite{snowflake}.
Apparently, a solid implementation of such a service within a service worker, capable of providing
strong anonymity guarantees (e.g., similar or close to Tor), is not
a trivial task and requires deeper analysis. Hence, such an exploration is beyond the scope of this paper.

\subsubsection{Click Fraud}
Having a large botnet can become profitable in many ways. One such way is to
abuse the digital advertising ecosystem, by having bots rather
than humans view or click on online advertisements.
It is estimated that online advertising fraud will cost advertisers \$19
billion in 2018, which represents 9\% of total digital advertising
spend~\cite{adfraud}.

\name can be easily used to generate clicks, as well as surf targeted websites
for a period of time, stream online videos to increase views, manipulate online
polls, and possibly sign up for newsletters.
To achieve that, the service worker can obtain periodically a list of online links
that is requested to visit, possibly combined with metadata such as visit duration, number of clicks, etc.
In addition, due to the rich programming features that \js\ offers,
\name can be easily programmed to follow a human-centric online behavioral model (e.g., similar to the one proposed by Baldi et al.~\cite{baldi2003}) to
evade countermeasures that seek to block users with unusual activity (e.g., clicking too many links in a short period of time).

\section{Evaluation}
\label{sec:evaluation}

\subsection{Prototype Setup}

To assess the feasibility and effectiveness of our approach, 
and also to check the existence of possible code protection and restriction mechanisms, we build a real world deployment of 
our \name prototype. 
Our prototype consists of two servers;
the first server is an Apache web server that hosts a simple webpage, and the second one is a command and control server (i.e., Puppeteer), delivering tasks to the Servants.
Upon the first website visit, the webpage registers a service worker in the Servant
and a sync manager that is 
responsible to keep the service worker alive in the background. 
After its registration, the Servant opens a full-duplex connection---using the WebSocket API~\cite{websocket}---with the Puppeteer and retrieves a 
\js\ code snippet that executes through \texttt{eval()}. In order to be able to use \texttt{eval()} from within the 
service worker, the collaborating web server gives the needed permission through the HTTP Content Security Policy (CSP).

\noindent{\bf Browser Compatibility:} As discussed in Section~\ref{sec:design}, our approach is based on existing components of HTML5 such as Service Workers and its
interface SyncManager.
Table~\ref{tbl:compatibility} summarizes the browser compatibility
of these components, and thus the compatibility of our framework.
As we can see, some vendors like Google, started supporting
service workers quite early (2016), while others caught up only until
recently, i.e., Safari (2018).
Still, \name is compatible with the most popular
browsers in both desktop and mobiles.

In our experiments we tested \name with four popular desktop browsers,
namely Chrome, Firefox, Opera and Safari.
However, we chose to exclude Safari from the performance evaluation results,
due to its bad performance sustained in all the experiments conducted.
Even though the service worker functionality is provided by Safari, we
experienced several performance glitches.
We believe that this behavior is due to the recently adaptation of service 
workers in Safari (2018).
Even for simple workloads, i.e., a simple counting example, the performance
achieved by the service worker is extremely slow (i.e., $20-50\times$ lower)
compared to the performance achieved by the
other three browsers.

\begin{table}[t]
	\caption{\name's browser compatibility}
	\label{tbl:compatibility}
	
	\centering
	\begin{tabular}{c|c|c}
		Device & Browser & SW compatibility\\ \toprule
		\multirow{6}{*}{Desktop} & Chrome & since v40 \\
		&Firefox & since v44 \\
		&Opera & since v26 \\
		&Edge & since v17 \\
		&Safari & since v11.1 \\
		& IE & NoSupport \\ 
		\midrule
		\multirow{10}{*}{Mobile} & Samsung Internet& since v4 \\
		& Chrome Android & since v64 \\
		& UC Browser & since v11.8 \\
		& iOS Safari & since v11.3 \\
		& Firefox Android & since v57 \\
		& Android Browser & Partially since v62 \\
		& Opera Mobile & Partially since v37 \\
		& Opera Mini & NoSupport \\
		& Blackberry & NoSupport \\
		\bottomrule
	\end{tabular}
\end{table}

\begin{figure}[t]
	\centering
	\includegraphics[width=0.95\columnwidth]{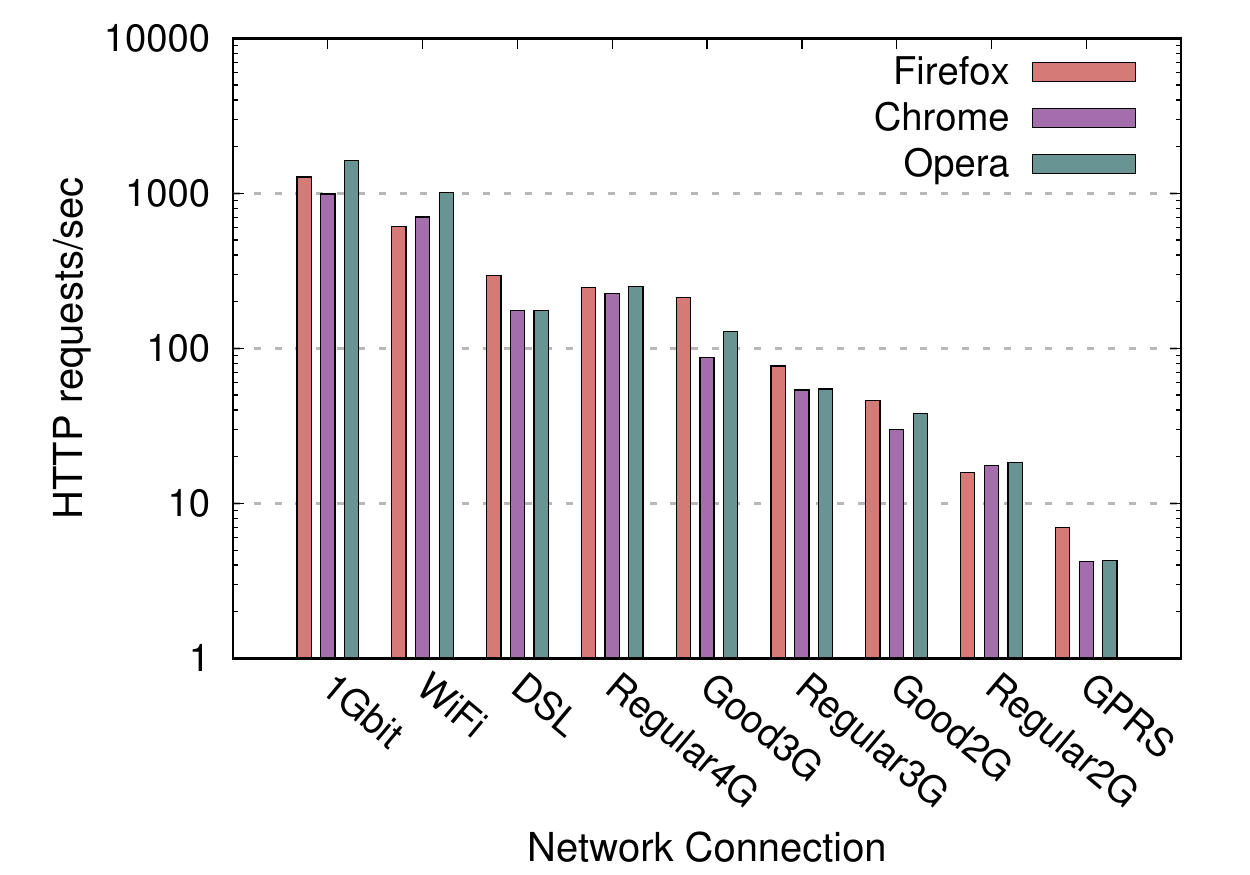}
	\caption{Rate of asynchronous outgoing HTTP OPTION requests for different browsers and network connections in the DDoS scenario. An orchestrated DoS attack in \name\ can achieve rates of up to 1632 reqs/sec per infected device.}
	\label{fig:ddos}
\end{figure}

\begin{figure*}[t]
	\centering
	\begin{minipage}{0.325\textwidth}
		\centering
		\includegraphics[width=1.05\columnwidth]{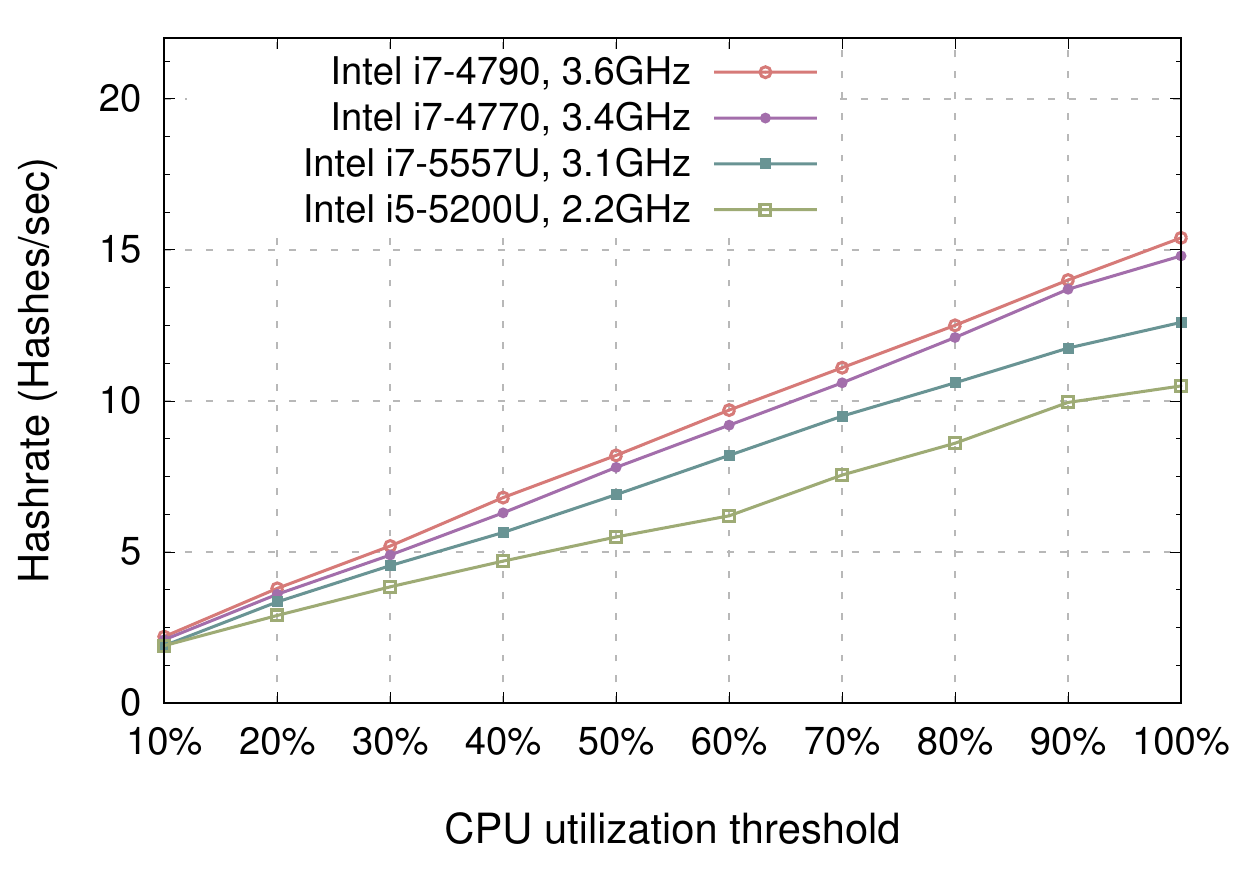}
		\caption{Hashrate for different equipped CPUs and utilization levels in the cryptojacking scenario. As expected, victim's hardware  
			affects significantly the computation power that \name\ may obtain.}
		\label{fig:hashrate_cpus}
	\end{minipage}
	\hfill
	\begin{minipage}{0.325\textwidth}
		\centering
		\includegraphics[width=1.05\columnwidth]{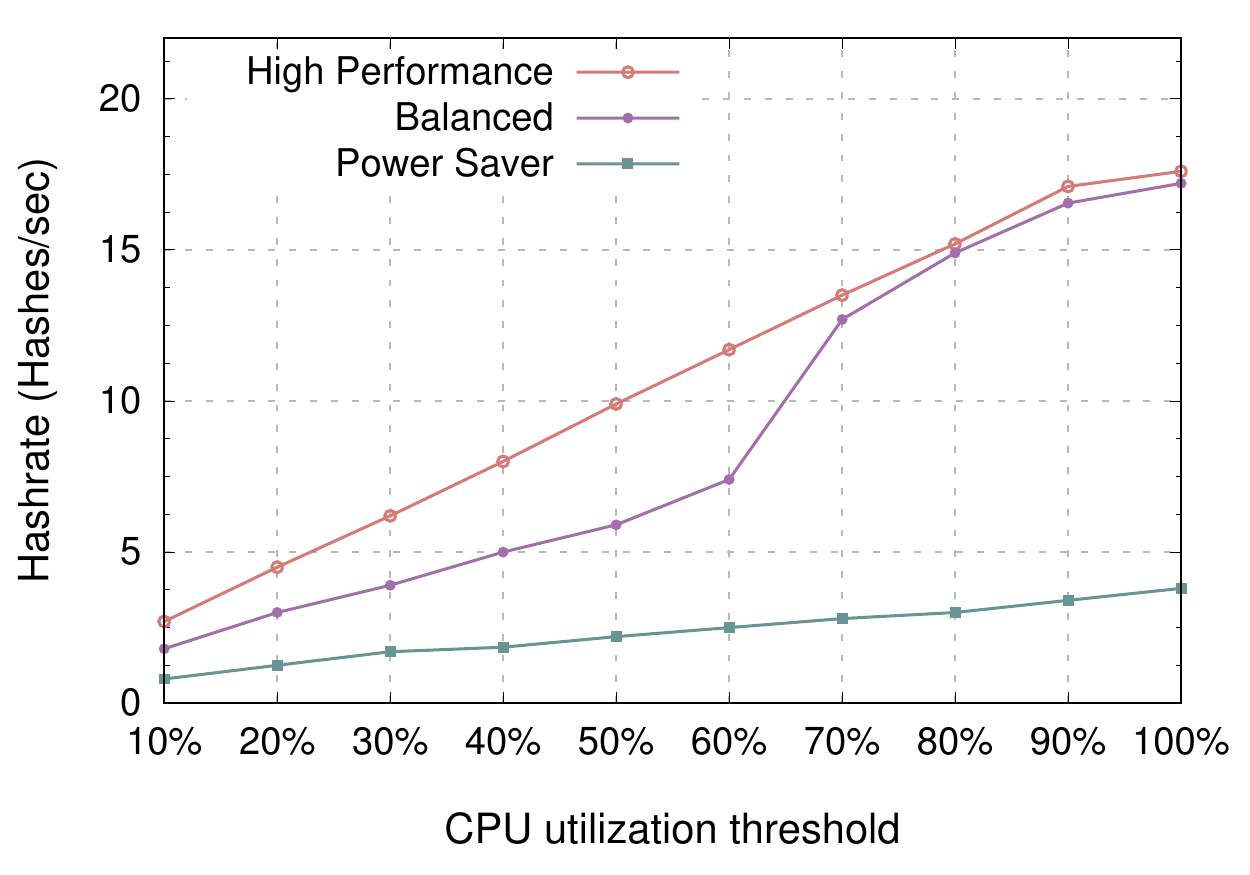}
		\caption{Hashrate for different utilization levels and Power modes in the cryptojacking scenario. The OS may slow down clock speed of the victim's device, reducing up to 78.41\% the computation power.}
		\label{fig:hashrate_win}
	\end{minipage}
	\hfill
	\begin{minipage}{0.325\textwidth}
		\centering
		\includegraphics[width=1.05\columnwidth]{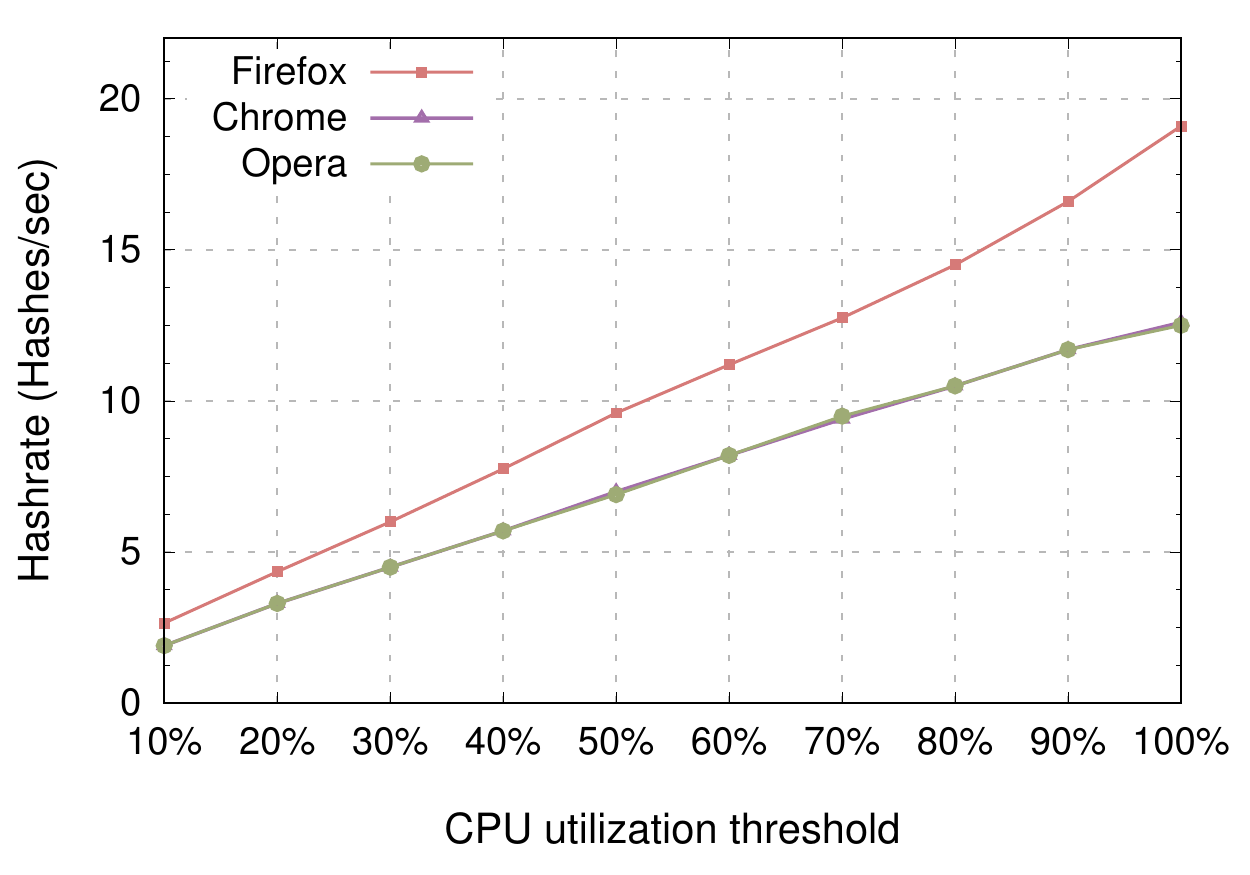}
		\caption{Hashrate for different infected browsers and utilization levels in the cryptojacking scenario. Firefox browser can calculate up to 
			$1.51\times$ more hashes per second than Chrome and Opera.}
		\label{fig:hashrate_browsers}
	\end{minipage}
\end{figure*}

\subsection{Performance Evaluation}

In order to demonstrate the effectiveness of \name, we conduct 
several experiments with various popular browsers and hardware settings, 
which allow us to make useful and interesting comparisons. 
However, it is noted that in this paper we do not aim to 
	provide an optimal implementation in terms of performance, 
	but rather to demonstrate the feasibility of the aforementioned attacks. To that end, the performance of the system can be further improved by using WebAssembly.
Furthermore, all the experiments presented in this section were conducted in a controlled environment, without reaching any host outside our local network (see Section~\ref{sec:discussion}).

\subsubsection{Abuse of network resources}

In the first experiment, we measure the rate of HTTP requests that the \name\ framework can achieve from a single browser.
As described
in Section~\ref{sec:ddos}, the Puppeteer 
instructs the Servant to continuously send multiple HTTP requests
to a remote server, via \texttt{XMLHttpRequest.send()}.
%
Figure~\ref{fig:ddos} shows the rate achieved for different browsers and different types of networks.
To measure the rate, we ran {\tt tcpdump} at the targeted server and
captured all the incoming HTTP traffic. 
As can be seen in Figure~\ref{fig:ddos}, even devices over inferior network connections are
capable of contributing a fair share in such a distributed
attack (e.g., an average of 214  reqs/sec on Good3G networks).
For high network bandwidth, i.e., 1 GbE, Opera tends to achieve higher rates (up to 1632 reqs/sec on average).

\subsubsection{Abuse of computation power}

The next experiment explores the computation 
capacity that the infected browsers can provide.
Figure~\ref{fig:hashrate_cpus} presents the hashrate achieved when mining Monero coins in Chrome,
for different CPU models and various utilization thresholds.
As expected, the performance gain is highly affected by the equipped hardware.
Specifically, we see that Intel i7-4790 can give 29\% more hashes 
per second than Intel i5-5200U, when fully utilized.

After experimenting with different operating systems, we noticed that the different power mode characteristics they provide can drastically affect the sustained performance of CryptoNight
execution.
Figure~\ref{fig:hashrate_win} shows the performance achieved on a Windows 7 desktop computer that is equipped with an Intel i7-4790K, at 4.0GHz, under 3 different
power modes (namely High Performance, Balanced and Power Saver).
When fully utilized, the Power Saver 
mode forces the CPU to reduce the voltage and clock speed, which causes a decrease of up to 78.41\% 
compared to the High Performance mode. In addition, in the Balanced mode, when CPU utilization exceeds 50\% 
the operating system allows the CPU to run in full speed in order to cover the increased computation needs, thus verging the 
hashrate of High Performance mode.

\begin{figure}[t]
	\centering
	\includegraphics[width=0.95\columnwidth]{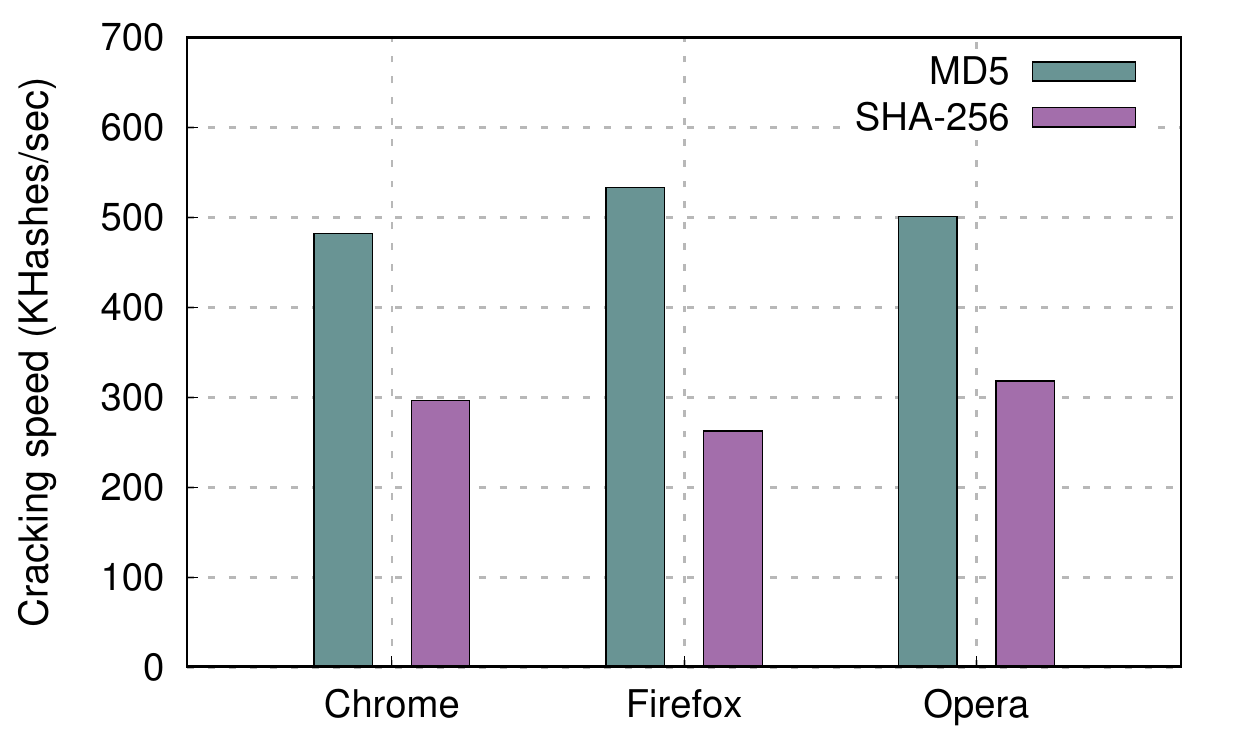}
	\caption{Cracking speed of different browsers in the distributed password-cracking scenario. \name\ can brute-force per victim around 500K MD5 hashes per second or around 300K SHA-256 hashes, irrespective of the infected browser.}
	\label{fig:password_hashing}
\end{figure}

Next, we set out to explore how different infected browsers affects the computation gain of \name.
Figure~\ref{fig:hashrate_browsers} shows the hash-rate achieved for different browsers, when
using a Intel i7-5557U CPU.
We observe that Firefox can calculate up to 
34.55\% more hashes per second than Chrome and Opera, which are both based on Chromium and the
V8 \js\ engine~\cite{chrome_opera}.

In our last experiment, we explore the performance sustained for password cracking.
Figure~\ref{fig:password_hashing} plots the achieved rate for
hashing 10-digit alphanumerical passwords on a brute-force manner, for both MD5 and SHA-256 algorithms.
As we can see, all browsers achieve similar 
and comparable performance. This means that a single browser can brute-force around 500K MD5 hashes per second or 300K SHA-256 hashes, 
irrespective of the infected browser. 

\subsubsection{Persistent and Evasive abuse}

In order to assess the persistence and evasiveness of our approach, we deliver \name\ within a webpage destined to perform cryptojacking. Before fetching the webpage in a Chrome browser, (i) we open {\tt tcpdump} and 
(ii) we deploy in our browser the following extensions/tools: Tamper Chrome HTTP capturing extension~\cite{tamperData}, Chrome's default DevTools,  WebSniffer~\cite{websniffer}, and HTTP Spy~\cite{httpspy} to explore in the real world, the  stealthiness of \name against state-of-the-art monitoring and blocking extensions.
After fully rendering the webpage and planting the Servant, we close the associated browser tab. Then, from the Puppeteer, 
(iii) we push a cryptocurrency mining task to the Servant and let it run for 3 consecutive days. 
We see that although the Servant regularly communicated with the Puppeteer to obtain PoW tasks, as {\tt tcpdump} correctly captured, 
\emph{none} of the employed extensions was able to monitor \emph{any} Servant-related traffic other than the very first GET request of the webpage, right before infection.

\noindent
\textbf{Comparison to state-of-the-art web botnets:}
In order to compare \name\ with the state-of-the-art web-botnets, we load our password cracking algorithm in a set of web workers 
as described in related approaches~\cite{dorsey,pan2016assessing}. 
Given that these web-botnets run only for as long as the victim is surfing the webpage, they need to fully utilize 
the resources of the infected device in order to scrounge a meaningful gain from this short infection window.
As a consequence, they usually occupy concurrently all system cores across the entire period of a website visit, which studies 
have shown that it is 1 minute on average~\cite{visitduration}. 

\begin{figure}[t]
	\centering
	\includegraphics[width=0.95\columnwidth]{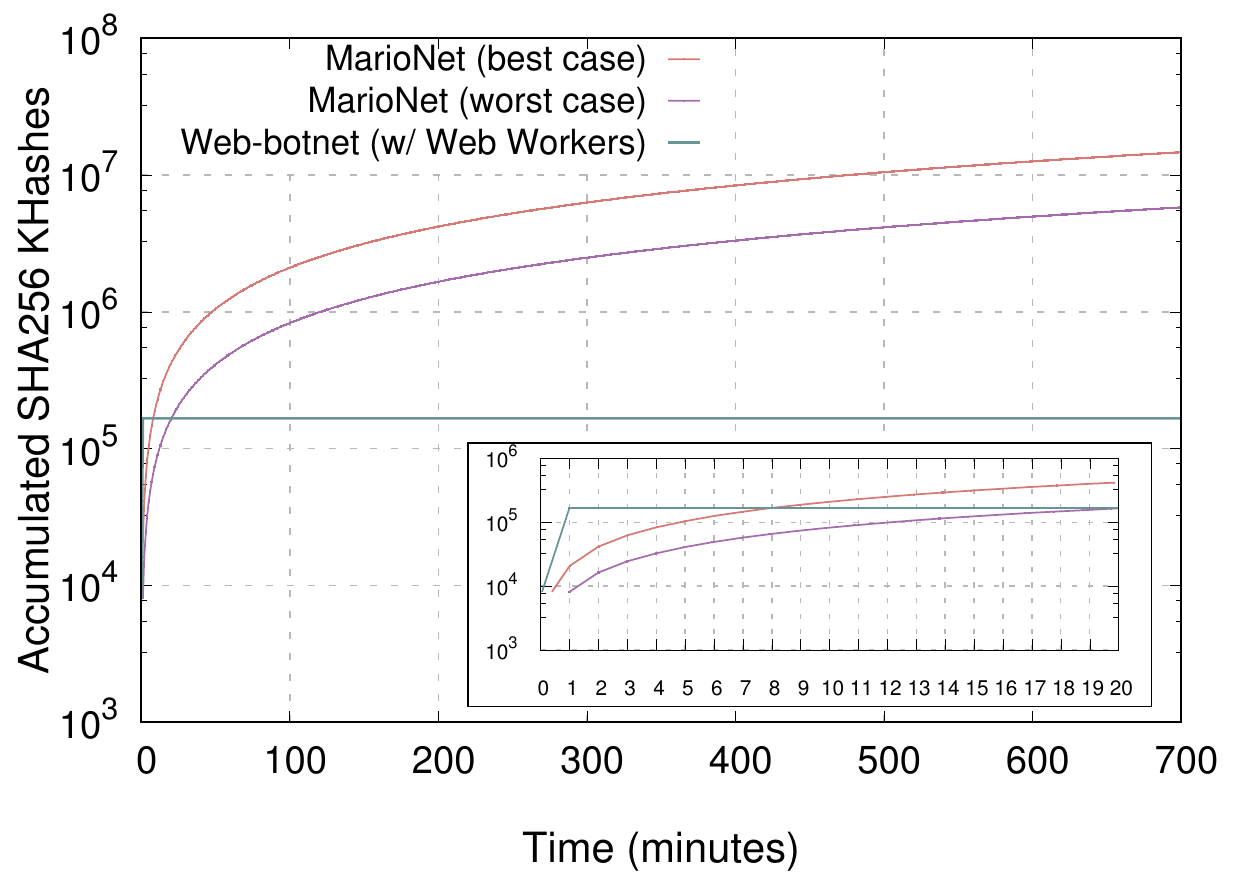}
	\caption{Number of SHA256 hashes brute-forced by \name and previous Web-botnet approaches that utilize Web Workers, in
		the lapse of time. The persistence of \name makes a 
		single infected browser compute hashes as long as the browser is open and thus be more efficient than opportunistic Web-Botnets.}
	\label{fig:comparison}
\end{figure}

In Figure~\ref{fig:comparison}, we plot the total number of SHA-256
hashes brute-forced by the 2 approaches in an infected browser for a period of 12 hours. For \name, we measure 
two cases: (i) the best case, where the password cracker runs uninterruptedly in the victim's device, and (ii) the 
worst case, where along with the malicious computations there is heavy utilization from other processes too. In the 
second case, to simulate this heavy load, we concurrently run a multi-threaded pi digit calculator~\cite{y-cruncher} 
that fully utilizes all 8 system's cores. As we can see, although web-botnet utilizes greedily $8\times$ more resources,
\name\ due to its persistence, enjoys a higher efficiency after the 18th minute of an open browser, even under extreme 
heavy concurrent interference.
Consequently, while till today, the business model of malicious websites were to deploy a web-botnet and find a way to keep 
the user on the website (by providing free movie streaming, online games or include pop-under windows~\cite{popunder}), 
with \name it takes only a momentary visit to infect the user and take control of their browser.

\section{Defenses} 
\label{sec:defenses}

In this section we examine potential defense mechanism that could detect and mitigate \name type of attacks.
The goal is to determine whether it is feasible to detect
the general methodology of the attack vectors that are opened through
the misuse of the service worker mechanisms, rather than
mitigating the specific use cases studied in this paper. 

We present various defense strategies and discuss the corresponding
tradeoffs they bring.
We categorize the defenses in two classes: (i) those that can be deployed
inside a vanilla browser (via the extensions/plugin mechanisms they support),
and (ii) those that can be deployed in the host (through anti-virus tools,
IDS/IPS, firewalls, etc.) or by modifying the browser.

\subsection{In-browser Mechanisms}

\subsubsection{Restricting Service Workers}

Service workers have been introduced to enable rich offline experiences, periodic
background synchronization, and push notifications among others.
Traditionally such functionality was normally requiring a native application.
However, the window of opportunity that open for attackers should make
questionable the tradeoff between rich user experience and security.

In the case of \name, disabling service workers could indeed prevent the
persistence and stealthiness of in-browser malicious computations.
However the capabilities service workers have, and the 
demand for such functionalities,
would make such a proposition inapplicable.
It would be feasible though to selectively enable service workers only
for some ``trusted'' websites, possibly via a browser extension that prevents/blocks registration of service workers at first place.

Another option would be to restrict their liveness, and be proportionate
to the user presence in the website that hosts them (i.e., the service
worker will be suspended after the user leaves the website) or apply a time cap (i.e., service worker gets terminated if it keeps running for an unreasonable amount of time).
By doing so,
the persistence characteristic of our attack will no longer be available.
The service workers will still be able to intercept and 
modify navigation and resource requests, as well as cache resources,
using the storage API for example, to allow applications to run even
when the network is not available.
However, service worker API is designed to provide important functionality to long-running web applications (e.g., Google Docs, email), allowing them to have rich offline experience. By forcing restrictions, service workers will not be able to provide
background synchronization (e.g., update the caches even when the user is
not at the site) or provide push messages and notifications.
Still though, a malicious actor could still be able to utilize them
in order to increase the stealthiness and evade detection by traditional
security browser extensions---i.e., a malicious computation taking place
inside a service worker will be more difficult to be detected.

A step towards striking stealthiness is to disable the \texttt{eval()}
family functions. By doing so, the attacker would need to ship the
malicious functionalities together with the service worker, which would
facilitate the signature-based content filtering browser extensions to
detect them easier. Obviously, this would be an arm race between attackers
and defenders, given the obfuscation and code scrambling techniques that
are in use for similar cases. In addition, service workers can include
minimal ISA emulators in order to execute malicious instructions received
from the attacker.
A more aggressive option would be to limit the functionality offered by
service workers, by making only a subset of \js\ available for
use (e.g., allowing only the sending/receiving of data between the
website and the web server). Clearly, such a data-driven approach would
require much more careful consideration and design.

\subsubsection{Whitelists/Blacklists}

Another possible defense strategy is to restrict the browser, with fine-grained policies, from fetching and deploying service workers. 
The simplest approach is the use of whitelists; i.e., service workers will be
blocked, unless the domain of origin is whitelisted.
These lists can initially include popular sites, which are typically considered more trusted, and enriched by web crawling and analysis platforms,
such as~\cite{provos2007}, that perform web-wide analysis to detect malicious
websites and \js\ files. 

\subsubsection{Click to Activate}

Another mitigation would be to require the user's permission for
registration and activation of a service worker---similar to
``Click to Play'' mechanism~\cite{clickToactivate} that disables by default plug-ins, such as Flash,
Java, Silverlight and others.
By doing so, the service worker functionality will be disabled by default,
and users would need to explicitly give permission for the service worker
to run.
Currently, this user consent is needed only for the Push Notifications~\cite{push}.
However, given the variation of attack vectors that can be achieved
through a malicious service worker, we believe that explicit permission would
raise user suspicion---in the same way it does for location, microphone,
etc.---especially when browsing unreliable websites.
One may say that these proposed permission-based defences may be not practical to constitute the perfect mitigation for the presented attack. However, recent developments such as GDPR, mobile or browser permissions model, etc. have demonstrated that user consent can be forced.

\subsection{Host-based Approaches}

\subsubsection{Signature-based Detection}

Traditional tools---such as firewalls, anti-virus, intrusion detection/prevention
systems--- are always a prominent methodology for the detection of malicious
activities.
The majority of these tools are typically using signatures to detect suspicious
data and/or code that enter or leave the user's computer.
The creation of such signatures for the case of \name may be trivial for some
attack cases.
For example, it could be easy to detect \name messages that are exchanged
between the service worker that lies in the browser and the back-end server, by
monitoring the network traffic.
A sophisticated attacker can obviously employ several techniques to raise the
bar against signature-based detection mechanisms.
For example, by installing end-to-end encryption with the back-end server can
sufficiently hide the content of the messages.
Given that a host-based approach can have full control of the client side though, the
SSL connection can be intercepted to acquire the decrypted data.
Besides, the messages
still need to be transferred, which can be a good hint for detection mechanisms
that are based on network flow statistics (e.g., number of packets exchanged, packet
size). Even though
covert channels and steganography may potentially help attackers, 
there are works that try to detect web-based botnets, by performing
anomaly detection on features like communication patterns and payload size~\cite{trafficFeatures}.


\subsubsection{Behavioral Analysis and Anomaly Detection}
A more drastic solution would be to develop techniques that
try to detect suspicious behavior of \js\ programs that are embedded in the web site or the
service worker.
Obviously, this would require more sophisticated analysis than
simple fixed string searching and regular expression matching, due to the fact
that the obfuscation of the malicious \js\ code snippets can evade static
analysis techniques. Instead, more advanced and complex analyzers should be used, such
as the monitoring of the utilized resources or the behavioral analysis of the executed code.
Even though this can be quite challenging, several works have been proposed in
the past~\cite{Hallaraker2005,Raman2008JaSPInJB,Heiderich2011}.
For instance, one of the first
anomaly detection approaches is JaSPIn~\cite{Raman2008JaSPInJB}, which
creates a profile of the application usage of \js\ and enforces it later.
IceShield~\cite{Heiderich2011} uses
a linear decision function that differentiates malicious code
from normal code based on heuristics for several attack types
that apply code obfuscation.
Finally, in~\cite{Hallaraker2005} the authors audit the execution of
\js\ code, and compare it to
high-level policies, in order to detect malicious code behavior.
Although all these approaches are not trivial,
they are a prominent step towards protecting against
malicious \js\ programs in general.

\section{Discussion}
\label{sec:discussion}

\noindent{\bf Ethical considerations.} In this paper, we implemented and deployed \name 
in a strictly controlled environment. During our experimentation 
with attack scenarios, no user or web server outside this controlled 
environment were contacted or 
harmed in any way. As such, we constrained the evaluation of our 
system to a limited set of nodes, thus avoiding any attempt to measure 
our system on a larger scale, in the real world.

\noindent{\bf Registration of multiple service workers.} 
Service workers are associated with specific scopes  
during registration and each service worker can only control 
pages that fall under its scope. If more than 
one service workers are registered (while the user is navigating 
throughout a website), then the browser enables only the service 
worker with the broader scope (typically the service worker 
registered at the root domain). However, during our experimentation
we observed that a publisher can design its website on purpose so
that multiple service workers can be registered in 
non-overlapping scopes (i.e., in file paths at the same level of the URI).
As a consequence, this allows \name\ 
to have multiple Servants simultaneously active and utilize
them for running its malicious tasks in multiple threads.


\noindent{\bf Cross-origin service workers.} 
The cross-origin service worker (or foreign-fetch~\cite{foreignfetch}) 
is an experimental feature of Chrome 54, to enable registration
of third-party service workers.
The motivation behind that, is to enable developers to implement advanced functionalities, such as client-side caching of CDN-based third party content.
However, this feature broadens the threat model of \name-like approaches, enabling third-parties to misuse the service workers of the domains that include them.
Even though this feature was discontinued one year after its announcement~\cite{foreignfetch2}, mostly
due to applicability issues, it still shows that such new functionalities
should be considered carefully in terms of security, before being applied.

Towards this direction, the aim of this work is to increase the awareness of developers and browser vendors regarding the 
provided powerful (but also potentially risky) capabilities of modern HTML. This way, restricting policies will adequately secure the user-side environments of future web applications.

%

%

\section{Related Work}
\label{sec:related}

Web browsers are a core part of our everyday life, being the door
to the gigantic world of  the web.
As a result, they have become a valuable target for attackers, that
try to exploit them in many different ways.

For instance, several approaches try to abuse the rich features of modern web applications, in order to form web-based botnets, the existence of which has
seen a significant rise recently~\cite{webbasedbotnets}.
Provos et al.~\cite{provos2007} present the threat of web-based malware infections, in which the infected browsers
pull commands from a server controlled by the attacker.
Contrary to traditional botnet-like attacks, web-based malware does not
spread via remote exploitation but rather via web-based infection. 
In~\cite{puppetnets}, the authors craft malicious webpages where users
get infected upon visit. The attackers can then abuse users' browsers
to perform attacks like DDoS, worm propagation, and node reconnaissance.
Grossmann and Johansen~\cite{millionDollar} leverage ads to deliver
malicious \js\ to users, forcing browsers to establish connections
with a victim server, thus performing a DoS attack.
A major limitation of these approaches though, is that the corresponding malicious
\js\ snippets need to be embedded in the main webpage. As a result,
long-running operations would block the rendering procedure and execution of the web application, making it practical only for short-lived attacks.

To overcome this limitation, many approaches started recently to use web
workers---a feature that was introduced with HTML5.
Web workers run as separate threads, and thus being isolated from the
page's window.
This allows the parallel execution of operations, without affecting the
normal rendering of the web application, leading to the rise of more
advanced web-based botnets.
Kuppan~\cite{HTML5Attacks} demonstrate this ability of using web
workers to perform DDoS attacks.
Rushanan et al. in~\cite{rushanan2016malloryworker}, also use web workers to
perform stealthy computations on the user side and launch not only attacks
like DoS and resource depletion but also covert channel using CPU and
memory throttling.
Pellegrino et al.~\cite{191940} also present different techniques to
orchestrate web-based DoS attacks, by utilizing web workers among other
HTML5 features, and provide an economic analysis and costs of browser-based
botnets.
Similarly, Pan et al.~\cite{pan2016assessing} explore the possibility of using web workers for  
performing application-layer DDoS attacks, cryptocurrency mining and password cracking.
Their results show that
although DDoS attacks and password cracking are feasible and with comparable financial cost, cryptocurrency mining
is not profitable for the attacker given the limited time a user spends in a website.
Dorsey presented an in-browser botnet using web workers as well~\cite{dorsey}.
The user browser, after infection, participates in a swarm of bots
performing various malicious 
operations like DDoS attacks, torrent sharing, cryptocurrency mining, and distributed
hash cracking. To infect as many 
users as possible, Dorsey embedded his malware in a malicious advertisement
and let the ad network to 
distribute it to the users browsers.
Similar to \name, all the above approaches do not require any software
installation on the user side. However, the browser remains under the
control of the attacker only for the duration that the user is browsing
the malicious website, making it impractical for long-running 
botnet operations. Instead, \name provides persistence that allows the 
attacker to perform malicious computations for a period longer than a 
website visit.


%

Besides the crypto-mining and crypto-jacking attacks, in which a
website unintentionally hosts web-mining code snippets~\cite{showtime,ronaldo},
there are publishers that intentionally use mining
to monetize their websites.
Eskandari et al. analyze
the existing in-browser mining approaches and their
profitability~\cite{DBLP:journals/corr/abs-1803-02887}.
Similar to web-based botnets, in-browser miners maintain a
connection with a remote server to obtain PoW tasks and abuse web workers
to achieve the highest possible CPU utilization on the user side.
However, the short website visiting times make the profitability
of this approach questionable~\cite{webminingwaste}.
\name also uses crypto-jacking as a possible scenario,
however instead of web workers
we leverage service workers to enable an entity to gain much
higher profits due to the provided persistence.

Finally, several attacks are based on malicious browser extensions that a user 
downloads and deploys in the browser~\cite{liu2011botnet,DBLP:journals/corr/abs-1709-09577}.
For instance, Liu et al. propose a 
botnet framework that exploits the browser extension update 
mechanism to issue batch commands~\cite{liu2011botnet}.
By doing so, they are able to perform DDoS attacks, spam emails
and passwords sniffing.
Similarly, Perrotta et al. exploit the over-privileged capabilities
of browser extensions to check the effectiveness of botnet attacks in contemporary
desktop and mobile browsers~\cite{DBLP:journals/corr/abs-1709-09577}.
Their results show that different attacks are feasible in different browsers.
A major difference of these approaches with \name, is that all the above
approaches require the installation of software (i.e., browser
extension) on the user side.


\section{Conclusion}
\label{sec:conclusion}

In this work, we presented \name: a novel multi-attack framework to 
allow persistent and stealthy bot operation through web browsers. 
Contrary to traditional botnet-like approaches, our framework does not require any
installation of malicious software on the user side. Instead, it leverages the existing technologies and capabilities provided by HTML5 APIs of contemporary browsers.

We demonstrate the effectiveness of this system by designing a large set 
of attack scenarios where the user's system resources are abused to perform malicious actions including DDoS attacks to remote targets, cryptojacking, malicious/illegal data hosting, and darknet deployment.
Two important characteristics of \name, that further highlight the severity
of the aforementioned attacks, is that it provides persistence, 
thus allowing an attacker to continue their malicious computation 
even after the user navigates away from the malicious website.
In addition, \name provides evasiveness, performing all operations in 
a completely stealthy way, thus bypassing the existing in-browser detection mechanisms.

Essentially, our work demonstrates that the trust model of web, 
which considers web publishers as trusted and allows them to execute code on the client-side without any restrictions is flawed and needs reconsideration. Furthermore, this work aims to 
increase the awareness regarding the powerful capabilities 
that modern browser APIs provide to attackers, and to initiate 
a serious discussion about implementing restrictions while 
offering such capabilities that can be easily abused.

\bibliographystyle{IEEEtranS}
\bibliography{paper}
\balance

\end{document}